4clean article first page with abstract and intro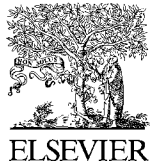

Available online at www.sciencedirect.com

ScienceDirect

Procedia Computer Science 00 (2012) 000–000

Procedia Computer Science

www.elsevier.com/locate/procedia
Conference title

# A comparison of SVM and RVM for Document Classification

Muhammad Rafi, Mohammad Shahid Shaikh

*FAST-NU, Karachi Campus, Karachi 75030, Pakistan***Abstract**

Document classification is a task of assigning a new unclassified document to one of the predefined set of classes. The content based document classification uses the content of the document with some weighting criteria to assign it to one of the predefined classes. It is a major task in library science, electronic document management systems and information sciences. This paper investigates document classification by using two different classification techniques (1) Support Vector Machine (SVM) and (2) Relevance Vector Machine (RVM). SVM is a supervised machine learning technique that can be used for classification task. In its basic form, SVM represents the instances of the data into space and tries to separate the distinct classes by a maximum possible wide gap (hyper plane) that separates the classes. On the other hand RVM uses probabilistic measure to define this separation space. RVM uses Bayesian inference to obtain succinct solution, thus RVM uses significantly fewer basis functions. Experimental studies on three standard text classification datasets reveal that although RVM takes more training time, its classification is much better as compared to SVM.

Keywords: Document classification, Support Vector Machine (SVM), Relevance Vector Machine(RVM)## 1. Introduction

Automatic method of managing large collection of documents, like clustering and classification, has received a lot of attention from research community. Without these robust and effective techniques we cannot cope with the rapid growth of document collections. Document classification is a task of assigning a new unclassified document to one of the predefined set of classes. The content based document classification uses the content of the document with some weighting criteria to assign it to one of the predefined classes. It is formally defined as, if there is a set of Documents D= {$d_1$, $d_2$ …….$d_n$}, where 'n' is a very large number. Now by using any text categorization algorithm, the document has to be semantically categorized in pre-defined set of classes or categories C= {$c_1$, $c_2$…..$c_k$}, where each $c_i$ represents a distinct class. There are numerous methods proposed in literature to perform the classification. The document classification task generally transforms the document into suitable compact representation for specific machine learning algorithm. It also performs pre-processing to enrich the representation like stop-word elimination, stemming and lemmatization. Traditionally, the documents become very compact after these pre-processing. Thus each word that remains in the representation is treated as a potential feature. Document is a feature vector of all these features in high-dimensional space. Many algebraic methods try to work for classification into this higher dimensional space. Classification methods [1,2] include nearest neighbour classifier [3], Neural networks [4], generative probabilistic classifiers [5] and support vector machine [6]. It is strongly agreed in the research community that support vector machine performance is exceptionally good on the task of document classification.

A new approach in SVM is to define a probabilistic vector on the same problem and it is termed as relevance vector machine, To the best of our knowledge it has not been applied for document classification task. In this paper, we investigate the performance of SVM and RVM learning methods on document classification task; the study has been carried out on standard datasets. In the realm of machine learning, a supervised approach to classification of documents takes place in three major steps: (i) Model training,( in our case the two models are SVM and RVM), (ii) Model validation on the test dataset (in our case 10-fold validation) and (iii) classification on unlabeled documents by model of SVM and RVM. The complete process is shown by the following schematic Figure 1.

Figure 1: Process of document classification

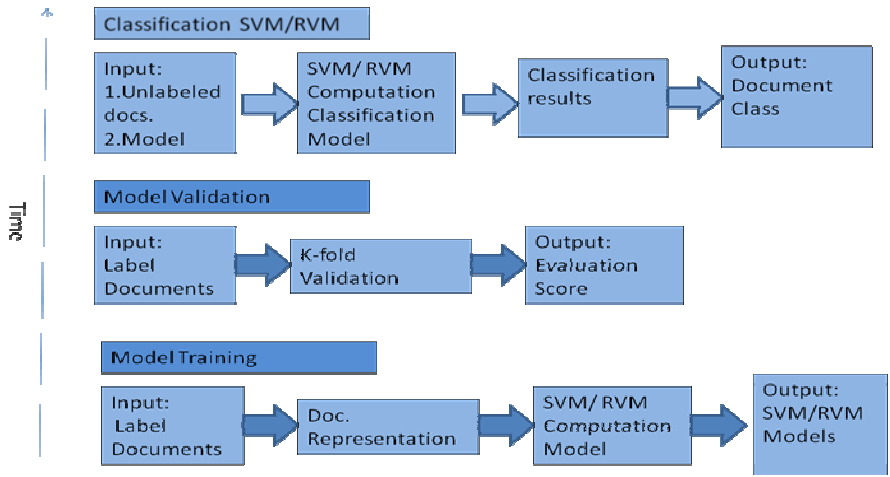

After testing the two approaches on three standard document classification datasets, we obtained results which clearly establish that RVM is much better in classification of documents. The rest of the paper is organized as follows: in the next section we discuss the recent work related to our study; secondly, we discuss document representation and description of support vector machine and relevance vector machine. The experimental setup and implementation is discussed in later part. Finally the results are discussed along with some useful direction to the future work.

## 2. State of the Art

Document classification is the process of assigning a given document into a predefined set of classes. The stable and rapid growth of document repositories poses new challenges in automatic classification. Generally, documents are transformed to a suitable representation before actually performing classification. There are numerous algorithms for the task of document classification. The rule-based method of classification of documents like [7,8] have limitation on generating a large number of rules and the classification process becomes a bit slow on large document collections.

Support vector machine (SVM) was first proposed in [9] for numerical data. The main idea of SVM is to determine separators in search space which best separate the different possible classes. The studies in [10] proved that SVM classification is very well suited for document classification as it favours sparse high-dimensional nature of textual data, which has very few irrelevant features. It is also suited for linear separable categories which are present in the documents. The very first classifier based on SVM that adopted for text/document classification is proposed in [10]. A comprehensive detail of this method is presented in [11]. It is also presented in [12] that SVM method is flexible and can easily be extended to perform interactive control on the task of classification. In our experiment, we have used multi-class SVM similar to the approach proposed in [13]. We have used "n" labelled training documents such that $\{(x_i, y_i)\}_{i=1}^{n}$ for $x$ here $x_i \in R^d$ which is standard vector representing for the i[th] training document. Each $y_j$ for j= { 1,2,3…K} is a unique class encoded as integer, hence there are total K such classes. On the other hand

Relevance Vector Machine (RVM) is based on the Bayesian formulation of a linear model with an appropriate prior. The main advantage of RVM, which was introduced in [14], is that it is more generalized as compared to SVM. RVM is sparser than SVM; similarly RVM is based on probabilistic prediction which can improve classification [15]. Considering these benefits of RVM we are motivated to use it on document classification tasks.

### 3. Document Representation

It is a standard practice in text mining that documents written in human language text are often preprocessed and transformed into a well-suited representation. Document representation is quite sensitive to a lot of text mining methods and document classification is no exception. In this study, we preprocess the three standard text mining datasets (see section 6). The documents are parsed; the stop-words and noise are removed. We also performed porter's stemming on each document. The resultant document-content is transformed into a suitable representation, which is later used for carrying out the classification task. Mathematically, a document $d_i$ which contains p distinct feature terms is represented as $R^p$ dimensional vector. $d_i = <w_{1i}, w_{2i}, w_{3i} \ldots w_{pi}>$, where the corpus space can be define as $D=<d_1, d_2, \ldots d_n>$.

### 4. Support Vector Machine (SVM)

Support Vector Machine (SVM) is a supervised classification technique in statistics and computer science. It is traditionally a two-class classifier. When given a set of input data (features), it tries to predict its respective class. Hence, it is a non-probabilistic binary linear classifier. SVM tries to build a model based on input set of training instances from the two-classes. This model can be used to classify new and unseen instances, given as input data to the model. The model predicts based on the generalized learning from the training data. SVM in it basic form represents the instances of the data into space and tries to separate the distinct classes by a maximum possible wide gap (hyper plane) that separates the classes, this instances mapped space is termed as model. The new query example for classification is again mapped to the model space; the query instance and its side-gap from the classified space determine the class nature of the query example. Mathematically, the SVM has a training data $D_T = \{(d_i, c_j) \mid d_i \in R^{di}, c_j \in (1,2..k)\}$.

Given a training document collection $D_T$, the label documents can be mathematically represented in high dimension space as

$$D_T = [\{(d_i, c_j) \mid d_i \in R^n, c_j \in (1,2,..k)\}] \quad Eq\ 1$$

$$C_i(W.d_i - b) \geq 1 \quad Eq-2$$

In Eq-2, Denotes the dot product, and W is the normal vector to the hyper plane. The parameter $\frac{b}{|W|}$ determines the offset of the hyper plane from the origin along the normal vector W. W and b must be chosen to maximize the margin. Thus for all classes Eq-2 should be

$$C_i(W.d_i - b) \geq 1 \quad \text{for all } 1 \leq i \leq K, \text{where } K \text{ is the number of classes}$$

$$\min_{W,b} \max_{\alpha \geq 0} \left( \frac{1}{2} |w|^2 - \sum_{i=1}^{n} \alpha_i [C_i(W.d_i - b) - 1)] \right) \quad Eq\text{-}3$$

We have used modified form of Eq-3 for performing SVM computation.

### 5. Relevance Vector Machine (RVM)

Relevance Vector Machine (RVM) is becoming a popular choice for classification task as it offers a number of advantages over SVM. First, it is based on Bayesian formulation of a linear model with prior that result in a sparse representation. It can be generalized very well and has low computational cost. Let the training data $D_T = \{(d_i, c_j) \mid d_i$ ε R$^{di}$, c$_j$ ε (1,2..k)}. The multi-class classification task on document can be modelled as RVM through the following mathematical model

$$c_i = W^T \times \varphi(d_i) \quad \textbf{Eq- 4}$$
$$c_i = \varphi(d_i, W) + \varepsilon_n$$

The final probabilistic classification equation of RVM based on Gaussian prior distribution of the document instance d$_i$ can be given as:

$$P(c_i/W, \sigma^2) = (2\pi \sigma^2)^{N/2} \exp(-1/2 \sigma^2 \times [c_i - \varphi(W)]^2 \quad \textbf{Eq-5}$$

The complete description of the derivation of the above can be found on [14]. We have used the above equation for computing the prediction of classes.

**6. Experimental Setup**

The experimental study is carried out by implementing SVM and RVM. The algorithms are implemented in Microsoft C# programming language. SVM and RVM are evaluated by performing a series of experiments on standard text classification datasets. We have used micro-average and macro-average F measure. For evaluating data set we have used 10 fold cross- validations and used paired t-test to assess the significance.

*6.1. Datasets*

REUTERS: The Reuters-21578, test collection of Distribution 1.0 is used. The collection appeared in Reuter's newswire in the year 1987. The collection consists of 22 data files, an SGML DTD file describing the format of the available data, and six files describing the categories used to index data. The collection is available at http://www.daviddlewis.com/resources/testcollections/reuters21578/

NEWS20: It is also a popular data set among text mining community; it's mainly used for text classification and clustering measure for machine learning techniques. The data set consists of approximately 20,000 newsgroup documents, partitioned in 20 different classes. The data set is available at http://people.csail.mit.edu/jrennie/20Newsgroups/

OHSUMED: The OHSUMED collection of the 1987-1991 abstract from 270 journals. It consists of over 348,566 references from the MEDLINE database, which is a database of medical literature maintained by the National Library of Medicine (NLM). Most of the references have abstracts and all have associated MeSH (Medical Subject Headings) indexing terms, with some of the MeSH terms marked as primary. The data set is available at http://davis.wpi.edu/xmdv/datasets/ohsumed.html

We have selected a subset of these datasets for our experimental studies. The following table gives a description of the selected subsets.

Table 1: Description of the datasets

| Dataset | Data Source   | Nos. Of Doc. | Nos. Of Classes |
|---------|---------------|--------------|-----------------|
| D1      | 20Newsgroup   | 200          | 7               |
| D2      | Reuters -21578 | 400         | 16              |
| D3      | OHSUMED       | 800          | 10              |
| D4      | 20Newsgroup   | 1600         | 16              |
| D5      | Reuters -21578 | 3200        | 20              |
| D6      | 20Newsgroup   | 3200         | 20              |
| D7      | Reuters -21578 | 6400        | 23              |

The datasets are preprocessed as per the standard practice of text mining. Each document is parsed and stop words are removed. After that stemming is performed using porter's algorithm.

*6.2. Micro/Macro Averaging F-measure*

The best method for measuring classification task is to calculate the percentage of correctly classified documents. F-measure is a measure of test accuracy. It is the harmonic mean of precision and recall. It is often used in information retrieval and text mining for measuring performance of classification. The Micro Averaging F-measure is the harmonic means of Micro precision and Micro recall. For obtaining this we calculate the True Positive (TP), False Positive (FP) and False Negative (FN) of the individual document independent of the dataset. It is a good instrument to measure the performance of classification method for individual document instances. Macro Averaging F-measure on the other hand, calculates the average precision and recall on individual datasets and takes average of all the datasets; it is an effective instrument when one wants to evaluate the algorithm on different datasets.

**7. Results and Discussion**

The paper successfully applied SVM and RVM for the document classification task. These models show good generalization capabilities. The results of Micro/Macro F-Measure on experimental study for classification of documents on standard datasets are reported in Table 2 below:

Table 2: Micro/Macro Average for the selected Datasets

| Datasets | SVM | | RVM | |
| --- | --- | --- | --- | --- |
| | Micro | Macro | Micro | Macro |
| D1 | 79.29 | 80.2 | 82.4 | 82.78 |
| D2 | 82.68 | 83.12 | 82.8 | 83.22 |
| D3 | 85.2 | 86.33 | 85.12 | 86.01 |
| D4 | 88.1 | 89.21 | 89.4 | 90.11 |
| D5 | 88.9 | 89.33 | 90.12 | 90.78 |
| D6 | 88.2 | 90.05 | 91.06 | 91.37 |
| D7 | 89 | 90.29 | 92.2 | 92.33 |

The result on Micro F-Measure for RVM is better than SVM, which is a clear indication that RVM can be used as a generalized document classification method. Similarly, for the Macro F- Measure RVM outperforms SVM on all the datasets except D3 (OHSUMED). A close inspection on the method revealed that this dataset consists of medical abstracts and the prior information for Bayesian inference is quite limited for short vocabulary. The result of dataset D7 (Reuters -21578) which contains large number of documents with arbitrary large number of classes clearly produced extra ordinary results for classification. The extra ordinary classification accuracy on RVM is mainly due to the fact that this dataset contains reasonably large text for building better prior probabilistic information. One good aspect of comparison for classification task in supervised techniques is the training time for the model. The following table gives time in seconds for building model for classification.

Table 3: Model Training Time in seconds

| Datasets | # of docs | TT-SVM | TT-RVM |
|---|---|---|---|
| D1 | 200 | 28 | 38 |
| D2 | 400 | 54 | 65 |
| D3 | 800 | 105 | 119 |
| D4 | 1600 | 198 | 241 |
| D5 | 3200 | 535 | 689 |
| D6 | 3200 | 578 | 688 |
| D7 | 6400 | 982 | 1082 |

The RVM training time is much higher than SVM as it computes prior information for predicting the class relationship which is not the case with SVM. When we compare the actual time for prediction the new example instances of document the two machines taking are almost the same time. Hence, we conclude that RVM takes more time in model building; however the actual prediction time is almost identical to that of a SVM. Similarly RVM model is more robust, contains minimum basis function and is more generalized. RVM performs good classification on the document datasets that contains reasonably large text contents.